# Approaching the absorption limit with monolayer semiconductor superlattices

Sara A. Elrafei[1], Lennart M. Heijnen[1], Rasmus H. Godiksen[1], Alberto G. Curto[1,2,3*]

[1] Department of Applied Physics and Eindhoven Hendrik Casimir Institute,
Eindhoven University of Technology, Eindhoven, The Netherlands

[2] Photonics Research Group, Ghent University-imec, Ghent, Belgium

[3] Center for Nano- and Biophotonics, Ghent University, Ghent, Belgium

* Corresponding author: Alberto.Curto@UGent.be

**Optical absorption plays a central role in optoelectronic and photonic technologies. Strongly absorbing materials are thus needed for efficient and miniaturized devices. There exists, however, a fundamental limit of 50% absorptance for any ultrathin film in a symmetric environment. Although deviating from these conditions allows higher absorption, finding the thinnest possible material with the highest intrinsic absorption is still desirable. Here, we demonstrate strong absorption approaching the fundamental limit by artificially stacking $WS_2$ monolayers into superlattices. We compare three simple approaches based on different spacer materials to surpass the record peak absorptance of $WS_2$ monolayers, which stands at 16% on ideal substrates. Through direct monolayer stacking without an intentional spacer, we reach a transmittance contrast of 30% for an artificial bilayer, although with limited control over interlayer distance. Using a molecular spacer via spin coating, we demonstrate controllable spacer thickness in a bilayer, reaching 28% transmittance contrast while increasing photoluminescence thanks to doping. Finally, we exploit atomic layer deposition of alumina spacers to boost the transmittance contrast to 36% for a 4-monolayer superlattice. Our results demonstrate that monolayer superlattices are a powerful platform directly applicable to improve exciton-polariton phenomena such as strong light-matter coupling and nanophotonic devices such as modulators and photodetectors.**

Strong absorption in ultrathin materials is sought after for light harvesting, photodetection, modulation, or sensing in photonic systems, both for free-space and guided-wave devices. There exists, however, a fundamental limit to the absorption of light by films that are much thinner than the wavelength: the ultimate achievable absorption is 50% when embedded in a symmetric refractive index environment.[1–3] Such a limit is a manifestation of the underlying distribution of electric dipole transitions: when a dipolar sheet is surrounded by a homogeneous medium, perfect absorption is not possible for incident light from a single direction due to symmetry. Nevertheless, it is possible to surpass this absorption limit by breaking the symmetry of the problem. One approach involves introducing a mirror positioned a quarter of a wavelength away from the thin layer. Such a Salisbury screen arrangement can lead to near-unity absorption.[3–6] Nanostructure arrays such as metasurfaces can also enhance local fields for increased interaction with a thin film.[7–10] Under interferometric excitation from both sides of the film, coherent perfect absorption is also possible under restricted conditions.[11] Despite the possibility of exceeding 50% absorption by deviating from the ultrathin-film limit conditions, there is a strong motivation to find stand-alone materials with higher intrinsic absorption, as they would be more universally applicable than extrinsic solutions.

Various thin-film materials can indeed provide strong and sharp excitonic absorption. Molecular J-aggregates of specific organic dye molecules can exhibit a narrow absorption peak in the visible or near infrared. Thanks to their strong oscillator strength, they have enabled reaching the strong coupling regime of light-matter interaction.[12–16] Despite their remarkable properties, J-aggregates still cannot reach the fundamental absorption limit as free-standing materials and lack versatility and tunability. Another promising class of strongly absorbing materials is two-dimensional crystals as they are easy to integrate into devices and allow electrical control. Adjacent monolayers are held together by van der Waals forces, which permits the isolation of atomically thin monolayers that retain the crystalline quality of the bulk material. Furthermore, the absence of surface dangling bonds facilitates stability in ambient conditions. As the bonding between monolayers and substrates occurs through van der Waals forces, lattice mismatch is not a limiting factor for creating heterostructures, unlike the case of conventional semiconductors grown epitaxially, such as Si or III-V materials. Graphene exhibits remarkably strong absorption for an atomically thin layer at mid-infrared to terahertz frequencies.[17,18] In contrast, its optical absorption is limited to 2.3% for a monolayer in the visible and near infrared,[18] and a large number of graphene layers is required for substantial absorption.[19,20] Monolayer semiconductors based on transition metal dichalcogenides such as $WS_2$ exhibit uniquely high absorption coefficients despite being less than one nanometer thick. They absorb 5-10% of broadband sunlight, an order of magnitude higher than GaAs or Si of comparable thickness.[21,22] Thanks to their strong and tunable excitonic resonances, they can serve as the basis for nanoscale optical components like metalenses, as highly reflective mirrors for spatial light modulators and active metasurfaces, or as photodetectors and modulators for integrated photonic circuits.[23–27] Their high oscillator strength and sharp exciton resonances are promising for reaching the strong light-matter coupling regime[7,28–30] and waveguiding using surface exciton-polaritons.[31,32] For such applications, where a strong excitonic resonance in a limited material volume is required, semiconductor monolayers are still suboptimal as they cannot reach the 50% absorption limit. As a solution, several monolayers can be stacked into superlattices with alternating monolayers and dielectric spacers. This promising route has been explored recently to enhance the absorption and photoluminescence of MOCVD-grown monolayers using a gold reflector.[31] Another recent study showed that bulk $MoS_2$ with monolayer properties can be obtained by molecular intercalation and dedoping.[33] Importantly, monolayer semiconductor stacks have been shown to improve the channel characteristics in stacked nanosheet transistors and could become relevant for industrial nano-electronics and optoelectronics.[34–36]

Here, we demonstrate three methods to create high-quality superlattices of exfoliated $WS_2$ monolayers to maximize ultrathin-film absorption. We focus first on structures containing two monolayers separated by a nanometric spacer to retain the optical properties of the constituent monolayers while keeping a minimal total thickness. We show enhanced absorption in artificial bilayers prepared by direct stamping, producing stacks without an intentional spacer. However, such structures are unstable, and further processing reduces the interlayer spacing, leading to decreased absorption. For increased robustness, we use a molecular spacer that also serves as a dopant, which yields increased photoluminescence as an added advantage. Finally, we produce higher-order superlattices by depositing spacers using atomic layer deposition. Although we demonstrate higher absorption in a 4-monolayer superlattice, the properties of the individual monolayers suffer degradation compared to the other fabrication methods, partially offsetting the scalability benefits. Our work demonstrates the potential of different monolayer superlattice configurations as a solid pathway to increase absorption using monolayer semiconductors for nanophotonic devices. As we use readily available fabrication methods, our results can be easily exploited in diverse applications requiring strongly absorbing ultrathin films, ranging from strong coupling physics to integrated photonics.

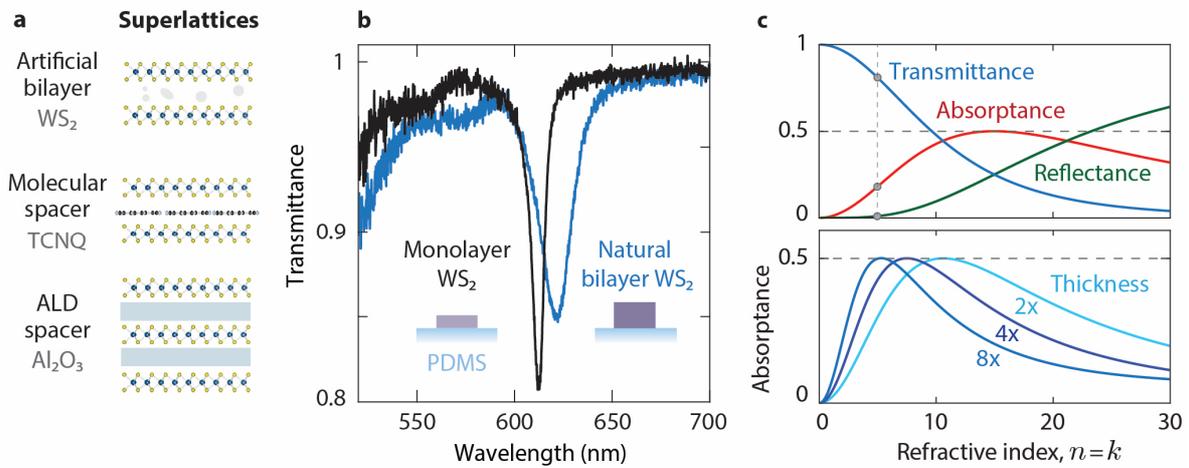

**Figure 1 | Stacking WS$_2$ monolayers in superlattices for higher absorption. a,** Approaches for stacking monolayers with different spacers: no intentional spacer, molecular spacer, and atomic layer deposition spacer. **b,** Transmittance of monolayer WS$_2$ on PDMS (black) showing a stronger and narrower exciton peak than an exfoliated bilayer (blue). **c,** Transfer matrix method calculations illustrating the ultrathin film absorption limit. Top: calculated transmittance, reflectance, and absorptance versus refractive index satisfying $n = k$, for a layer 0.62-nm thick in a symmetric PDMS environment ($n_{PDMS} = 1.42$) at $\lambda = 613$ nm. The refractive index of the WS$_2$ monolayer at the A-exciton peak is close to $5 + 5i$, indicated by the vertical dashed line. Bottom: absorptance for increasingly thicker films in multiples of the monolayer thickness, showing maximum absorption for lower refractive indices along the $n = k$ condition as the number of layers increases.

**The maximum absorption limit for monolayer semiconductors**

We chose monolayer WS$_2$ as the absorbing material due to its record-high absorption coefficient. We start by mechanically exfoliating WS$_2$ monolayers on a polydimethylsiloxane (PDMS) film attached to a glass slide. The PDMS substrate provides the best starting point known for WS$_2$ at room temperature:

a high peak absorption, narrow exciton linewidth, and high quantum efficiency.[37] Compared to alternative substrates such as SiO$_2$ on Si or glass, PDMS results in higher absorption and more uniform properties over the monolayer (Supporting Section S1). The transmittance spectrum of monolayer WS$_2$ on PDMS exhibits an exciton resonance with a transmittance contrast of 17% (Figure 1b and Supporting Section S2 for reflectance and absorptance). The main excitons in the transmission spectrum at wavelengths of 612.4 and 515.8 nm are referred to as the A and B excitons, respectively, along with a smaller dip due to the first excited state of the A exciton ($n$ = 2).[38] Although monolayer WS$_2$ already offers remarkable absorption, an ideal ultrathin material could still reach 50% absorptance.

As a first candidate for improved absorption over the monolayer case, one might consider using a bilayer crystal. We compare a monolayer and a natural bilayer in Figure 1b, where the natural bilayer exhibits a lower transmittance contrast of around 14% and a significantly broader linewidth than the monolayer (54 and 38 meV, respectively). The higher absorption and narrower linewidth in a monolayer compared to a bilayer can be attributed to the absence of interlayer coupling, the reduced dielectric screening, and the enhanced radiative rate of excitons in the monolayer.[5] These results indicate that natural bilayers or multilayers are not viable for maximizing absorption with narrow linewidth. Instead, we turn to artificial structures by purposefully stacking multiple monolayers into superlattices. We can maximize absorption in an ultrathin film with minimal thickness by maintaining a controllably low interlayer coupling yet with the shortest possible interlayer spacing. We will demonstrate that this approach effectively enhances absorption by preserving the direct bandgap character of the monolayers, thus facilitating the accumulation of a strong excitonic response.

We adopt the transfer matrix method to understand how light interacts with atomically thin semiconductors and their superlattices (Supporting Section S3). Given the thickness and complex refractive index of each layer, $\tilde{n} = n + i\, k$, we calculate the transmittance (*T*), absorptance (*A*), and reflectance (*R*) for thin-film stacks surrounded by two semi-infinite media under plane-wave illumination at normal incidence.[3] First, we investigate a 0.62-nm-thick film, corresponding to a single WS$_2$ monolayer but with variable refractive index, surrounded by PDMS (Figure 1c, top). We sweep the refractive index along equal real and imaginary parts of the refractive index ($n = k$), which is a necessary condition to reach the absorption limit for ultrathin films.[2,5,39] We find indeed an absorptance maximum reaching the 50% limit with transmittance and reflectance equal to 25%. Nevertheless, maximum absorption occurs when $n = k \approx 15$ for a thickness equivalent to a single monolayer, which is not feasible in realistic optical materials. Our WS$_2$ monolayers exhibit refractive indices close to $n = k \approx 5$ (Figure 1c, dashed vertical line). As the film thickness increases to multiples of a monolayer, the requirement for maximum absorption shifts to lower values of the refractive index (Figure 1c, bottom). We predict thus that it is possible to reach the absorption limit with a realistic WS$_2$ refractive index using a film thickness equivalent to 8 monolayers.

**Stacking monolayers into artificial bilayers**

As the first step towards superlattices with multiple monolayers, we investigate artificial bilayers consisting of two stacked monolayers. To fabricate them, we deposit a WS$_2$ monolayer directly on top of another monolayer based on a dry stamping method using PDMS (see Methods). The resulting artificial bilayer is encapsulated in PDMS (Figure 2a, inset). Van der Waals forces bind the stack together, and the interlayer space contains air and impurities introduced during the fabrication process.[40–43] Such an artificial bilayer shows a transmittance contrast of 30% and a slight increase in linewidth compared to the constituting monolayers (Figure 2a, top). When reflection is taken into account, we find an absorptance of 27% (Supporting Section S2). The photoluminescence intensity (PL) of the artificial bilayer roughly doubles the monolayer value (Figure 2a, bottom), which points to the possible absence of interlayer coupling as the origin of the preservation of the monolayer exciton quality in artificial bilayers. We observe a slight increase in linewidth and a small redshift, likely due to the higher permittivity of the surroundings of both monolayers due to the presence of the other monolayer, the additional PDMS superstrate, and impurities introduced in the stamping process.

We exploit hyperspectral imaging to quantify the spatial homogeneity of the exciton properties over an artificial bilayer. We record a spectrum at every point of a specified sample region to obtain statistics of several exciton properties in each area. Using hyperspectral transmission images, we retrieve the energy, linewidth, and transmittance contrast at the A-exciton peak (Supporting Section S4). There is a consistent rise in transmission contrast from 17% for a monolayer to 28% for an artificial bilayer. Furthermore, mapping shows the uniformity of the monolayer and artificial bilayer regions (Figure 2b), as confirmed by histograms with almost no overlap in transmittance contrast for the monolayer and artificial bilayer (Figure 2c). Additionally, we quantify a red shift in peak energy of approximately 1.5 meV.

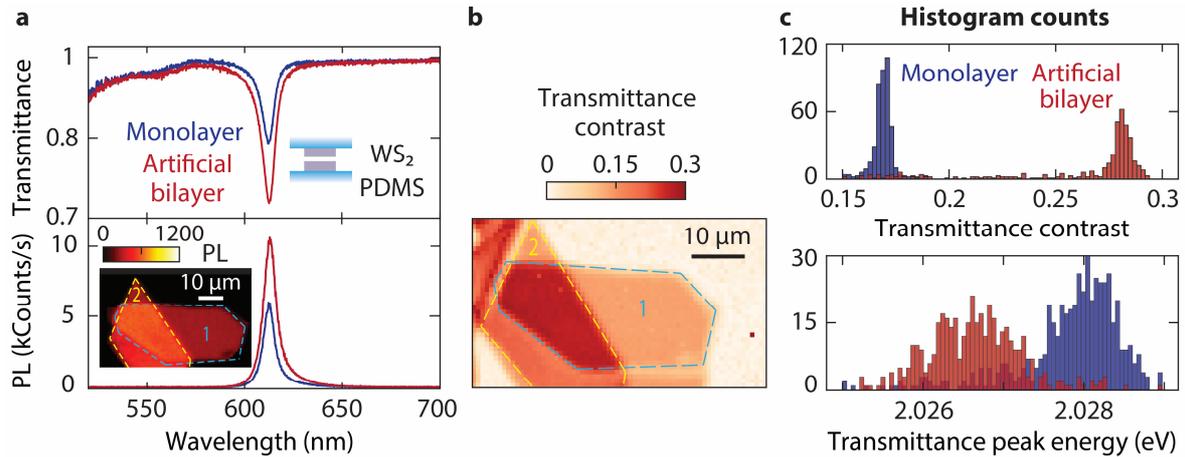

**Figure 2 | Increased absorption in an artificial WS$_2$ bilayer consisting of two stacked monolayers. a,** Comparison of WS$_2$ monolayer (blue) and an assembled bilayer (red) in a symmetric polymer environment. Inset: PL map with higher luminescence on the overlap area between Monolayers 1 and 2 (blue and yellow outlines). **b,** Hyperspectral image of the same bilayer using transmittance contrast at the A-exciton peak energy. **c,** Histograms obtained from hyperspectral transmittance imaging of monolayer and artificial bilayer areas. Top: transmittance contrast at the A-exciton peak. Bottom: A-exciton peak energy.

Although this method of producing bilayers is simple and already yields high absorption, we show next that the interlayer spacing is not fully reliable. We investigated the variability of the gap between the stacked monolayers. The stacks displayed consistent optical properties despite their random relative orientations, suggesting the presence of a small unintentional gap. This gap is likely filled with bubbles, exfoliation residue, and impurities from the transfer process. To reduce and control this spacer, we investigate the effect of applying heat and vacuum on a bilayer stack. Initially, the monolayers are nearly perfectly decoupled, doubling the PL intensity (Figure 3a, top). However, after heating the artificial bilayer at 80 °C for 20 minutes, interlayer interaction starts to appear, causing a PL gradient in the central area of the bilayer (Figure 3a, middle). As the temperature increases, the transmittance and PL spectra are slightly red-shifted while contrast diminishes. As the gap closes further, the PL intensity decreases significantly, indicating a transition from weak coupling to strong interaction leading to changes in the optical properties. After storage in vacuum, the spacer finally collapses (Figure 3a, bottom): the artificial bilayer starts resembling a natural bilayer with quenched PL and lower transmittance contrast accompanied by a broader linewidth (Figure 3b).

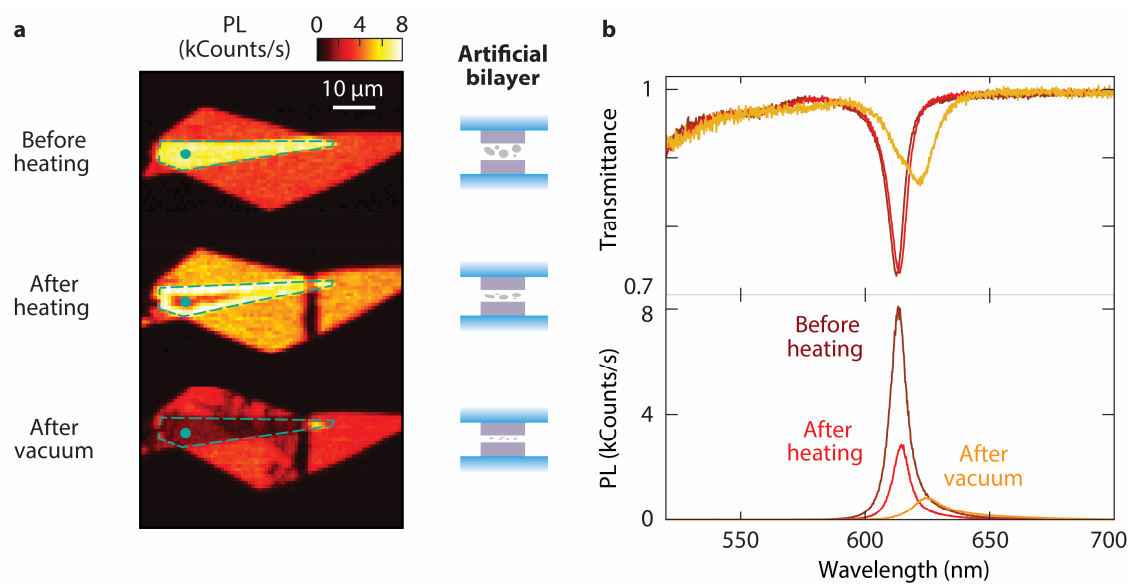

**Figure 3 | Control of spacer thickness in an artificial bilayer. a**, Confocal PL maps of stacked monolayers at different stages of closing the interlayer gap upon thermal annealing and vacuum storage. **b,** Corresponding transmittance and PL spectra. After such treatments, the artificial bilayer area is quenched with spectra similar to a natural bilayer.

**Molecular spacers between two monolayers**

To reliably maintain the desired monolayer properties by setting a minimum gap distance, we incorporate an intentional spacer layer. First, we employ a molecular spacer to control the interlayer coupling more precisely than in our previous artificial bilayers by introducing tetracyanoquinodimethane (TCNQ) molecules between the two monolayers. Interestingly, the adsorption of TCNQ molecules also induces charge transfer and provides *p*-type doping. It brings the *n*-type $WS_2$ monolayer closer to an intrinsic semiconductor and increases its PL quantum efficiency without significantly altering the exciton energy and linewidth.[44–47] We prepare the molecular spacer by spin coating a $WS_2$ monolayer on PDMS with TCNQ in methanol (see Methods). We focus first on a single monolayer and compare the transmittance and PL spectra of the monolayer with and without TCNQ doping. Using a TCNQ concentration of 1 mM, we observe an increase in the PL intensity of the monolayer, approximately three times higher than before doping (Figure 4a). As doping does not significantly change the transmission spectrum of a monolayer, these results verify the potential of TCNQ as an advantageous molecular spacer in superlattices.

We fabricate a bilayer with a molecular spacer by stamping an undoped monolayer on top of a doped monolayer. For a fair comparison with the previous artificial bilayers, we applied heating at 75°C for 15 minutes and vacuum to ensure better contact between the layers. The molecular spacer yields a bilayer transmittance contrast of 28% (Figure 4b), corresponding to approximately 25% absorptance (Supporting Section S2). In addition, it provides a 1.5-fold enhancement in PL intensity of the molecular spacer bilayer compared to a single doped monolayer. We attribute the less than twofold increase to the sharing of the molecular dopant layer between both monolayers. The monolayer and bilayer areas are highly homogeneous without internal PL features (Figure 4b, inset).

We also show the dependence of the exciton properties in a molecular spacer bilayer on the TCNQ concentration. Spin coating with different concentrations corresponds to different molecular layer thicknesses. Our results indicate that increasing the spacer thickness using higher concentration enhances the transmittance contrast and reduces the linewidth of the A exciton, which also shifts to the blue (Figure 4c). At a concentration of 0.1 mM, the monolayers interact strongly due to an insufficient spacer thickness, causing quenching and spectral broadening reminiscent of a natural bilayer. In summary, using a molecular spacer provides dual benefits for the optical properties of stacked monolayers: it enhances PL while facilitating improved light absorption. Despite the clear potential of molecular spacers in optimizing the optical properties of stacks of layered materials, our fabrication method cannot be easily scaled to thicker superlattices to further approach the ultrathin film absorption limit.

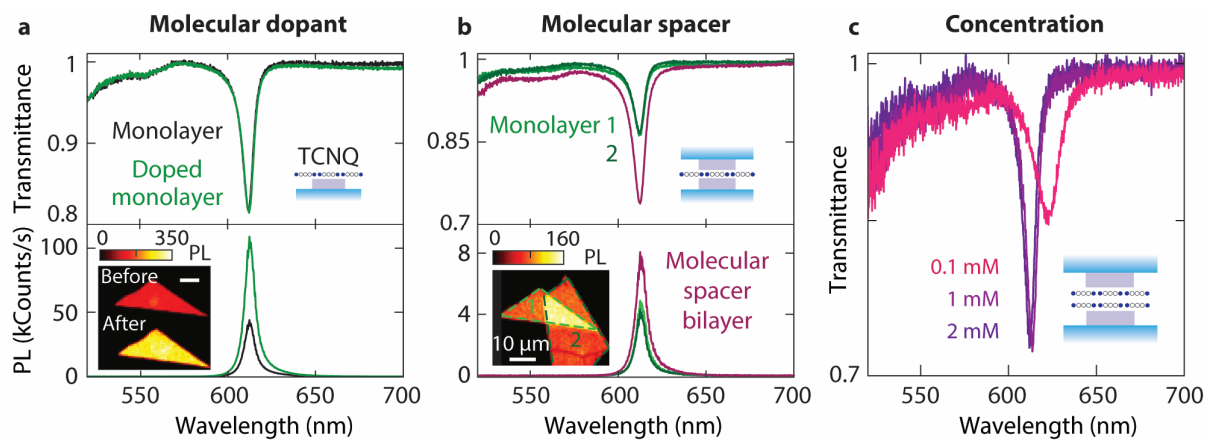

**Figure 4 | Molecular spacer bilayer with controlled interlayer distance**. **a,** A WS$_2$ monolayer with and without TCNQ doping: transmittance and PL spectra. Inset: PL maps illustrating the emission enhancement of a monolayer before and after doping with a TCNQ concentration of 1 mM. **b,** Transmittance and PL spectra of an assembled WS$_2$ bilayer with molecular spacer using a TCNQ concentration of 1 mM. Such a bilayer exhibits higher absorption and PL intensity than the monolayer. **c,** Effect of molecular spacer concentration on the transmittance contrast.

**Scalable superlattices through atomic layer deposition of spacer layers**

The artificial and molecular spacer bilayers benefitted from a PDMS substrate to maximize exciton absorption. Using a rigid substrate instead of a flexible polymer film typically results in lower monolayer absorption, but it also enables compatibility with other spacer deposition methods that facilitate the scalability of monolayer superlattices. Next, we use atomic layer deposition (ALD) to grow alumina (Al$_2$O$_3$) spacer layers (see Methods). We deposit 50 nm of Al$_2$O$_3$ on amorphous quartz as a substrate for the first WS$_2$ monolayer. We prepare superlattices using again the PDMS dry transfer method with subsequent spacer layers deposited using ALD. For a bilayer stack with a 2-nm-thick spacer, we observe a transmittance contrast of 23.9% (Figure 5a). We attribute the reduction compared to the previous fabrication methods to the presence of a different substrate, strain, and damage during the ALD process (Supporting Section S5).[37,48]

We fabricate thicker superlattices with 3 and 4 alternating WS$_2$ monolayers and spacers. The transmittance contrast increases with the number of layers, as seen in a hyperspectral transmittance image at the A-exciton peak (Figure 5b). The transmittance contrast for 1-, 2-, 3-, and 4-monolayer superlattices are 14.2, 23.9, 30.6, and 36.4%, respectively. The contrast for the B exciton to the blue of our spectral range also grows with the number of monolayers (Figure 5a). Relying solely on transmission as an indicator for absorption is insufficient because reflection cannot be neglected for structures with increased absorptance (Figure 1c and Supporting Section S2). Taking into account reflection, the absorptance of the 1-, 2-, 3-, and 4-monolayer superlattices are 13.8, 21.6, 26.6, and 30.8%, respectively.

The PL of a superlattice initially increases with the number of layers, but the peak intensity does not grow linearly for additional layers (Figure 5a, bottom). Although we kept a low deposition temperature of 100 °C to minimize damage, the additional defects due to the ALD process and trapped impurities introduced by stamping negatively affect the quantum efficiency of each monolayer cumulatively. Consequently, emission saturates for 3- and 4-monolayer superlattices (Figure 5a, inset). To conclude, building superlattices with ALD spacers provides a scalable route to increased absorption and brighter light emission, although the spacer deposition process needs further optimization to fully take advantage of this approach to reach the 50% limit with the thinnest possible superlattice.

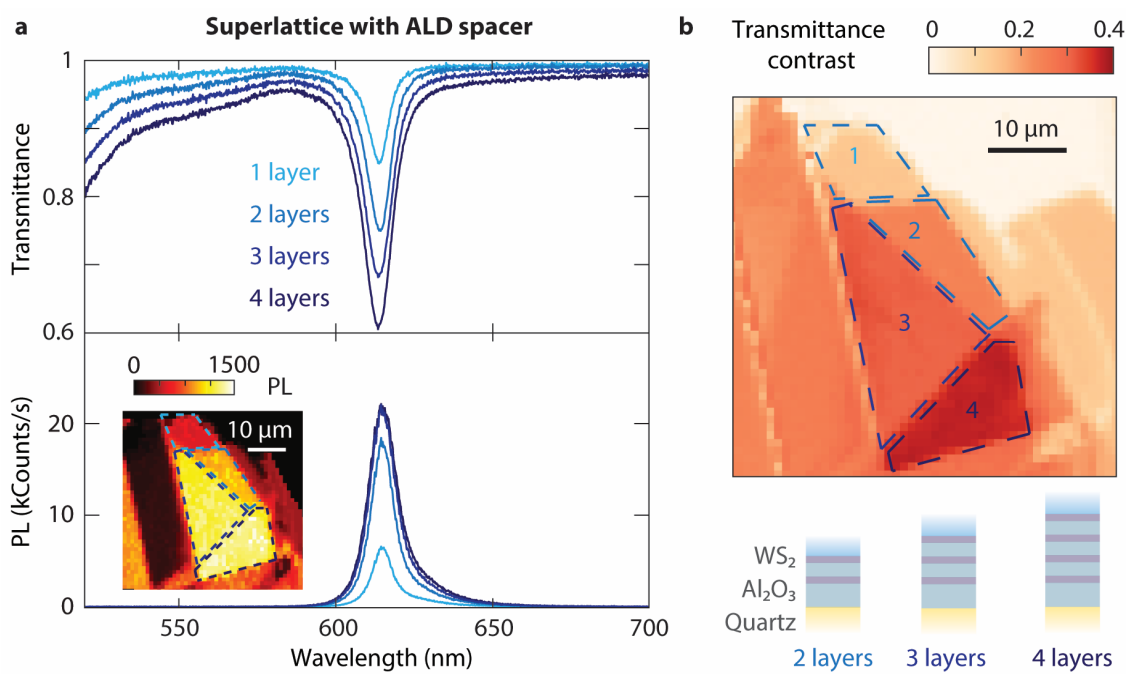

**Figure 5 | Superlattice with alternating WS$_2$ monolayer and Al$_2$O$_3$ layers deposited through atomic layer deposition. a,** Transmittance and PL spectra of superlattices with increasing number of WS$_2$/Al$_2$O$_3$ unit cells. **b,** Map of transmittance contrast variations for different superlattice thicknesses. Schematic illustration of the assembled layers for 2, 3, and 4 WS$_2$ monolayers. The stacks are protected with PDMS capping.

**Conclusion**

We demonstrated the efficacy of superlattices formed by stacking multiple $WS_2$ monolayers to maximize ultrathin film absorption. Superlattices provide higher absorption than high-quality exfoliated monolayers and bilayers while preserving the narrow exciton linewidth and emission quantum efficiency of single monolayers. We compared three approaches based on direct stacking without an intentional spacer, with a molecular spacer, and with an alumina spacer. Creating an artificial bilayer through simple stacking led to a substantial increase in transmittance contrast to 30% from 17% for a monolayer, albeit with limited control over gap thickness. Introducing a molecular spacer using TCNQ proved highly effective, offering controllable thickness through varying molecular concentrations. A bilayer stack with a molecular spacer allowed us to reach 28% transmittance contrast. Additionally, this molecular spacer also increases photoluminescence by acting as a dopant. Finally, we used atomic layer deposition of insulating spacers to increase the number of monolayers and boost the transmittance contrast to 36%. Further fabrication optimization and scaling of these monolayer superlattices is still possible.

The versatility of our superlattices and the simplicity of the fabrication methods present compelling advantages for their incorporation into devices that would benefit from improved absorption. Our demonstration opens avenues for nanophotonic devices with superior light harvesting, emission, or modulation capabilities. Moreover, monolayer superlattices can be leveraged to easily improve light-matter interaction in the strong coupling regime and waveguiding using surface exciton-polaritons compared to single monolayers. Beyond increased absorption, engineering superlattices to fine-tune the optical properties of semiconductor monolayers can enable nanoscale sensors and sub-nanometric rulers based on changes in exciton properties. Monolayer semiconductor superlattices offer thus solutions for efficient and miniaturized nano-optoelectronics and integrated photonics converging with future atomically thin electronics.

**Methods**

**Preparation of monolayer superlattices**

We mechanically exfoliate a bulk $WS_2$ crystal (*n*-type doping, HQ Graphene) down to a monolayer using tape (SPV 9205, Nitto Denko) onto an optically transparent PDMS film (Gel-Pak PF-80-X4). We identify monolayer areas using a fluorescence wide-field microscope and transfer them to target positions using two *XYZ* stages. The PDMS film acts as a stamp for all-dry viscoelastic stamping[49] used in the following three approaches to construct monolayer stacks.

*Artificial bilayer*

The PDMS film, with the first monolayer on top, lies on a glass slide. We transfer a second monolayer carried by another PDMS film onto a target monolayer on the PDMS/glass substrate. We leave the top PDMS film on the stack to prevent unnecessary damage to the monolayers. After initial optical characterization of the as-deposited bilayer, we treat some samples to test the robustness of the interlayer gap. Initially, we subjected those samples to thermal annealing at 80 °C for 20 minutes on a hotplate in air, to remove bubbles and impurities. Subsequently, we stored the sample in a vacuum desiccator with a VWR VP 86 diaphragm pump to ensure uniform layer-to-layer contact.

*Molecular spacer*

The molecular spacer approach starts with a monolayer on a PDMS film supported on a glass slide. To create the molecular spacer, we use 7,7,8,8-tetracyanoquinodimethane (TCNQ, Ossila Ltd). We prepare solutions with different molecular concentrations in methanol: 0.1, 1, and 2 mM. For example, we dissolve 4.1 mg of TCNQ powder in 20 mL of methanol for a concentration of 1 mM. To spin-coat a $WS_2$ monolayer on a PDMS/glass substrate, we pipette 20 μL of TCNQ solution followed by spinning for 1 minute at 500 rpm. At this speed, the deposition produces a thin, flat, and homogenous film[50–52] with minimal damage to the monolayer semiconductor. The TCNQ molecules are expected to lie flat on the monolayer, which facilitates their use as a controlled spacer. Next, we stamp a second monolayer on top of the TCNQ-coated monolayer. As in the case of the artificial bilayer, we include thermal annealing and vacuum treatment. We leave the PDMS superstrate covering the resulting molecular spacer bilayer.

*Atomic layer deposition spacer*

For the monolayer superlattices based on ALD spacers, we transfer the monolayers from PDMS onto ALD-coated, amorphous quartz wafer pieces (Planoptik). We deposit the $Al_2O_3$ layers using a high-vacuum ALD system (FlexAL2, Oxford Instruments). Trimethylaluminum (TMA) and water vapor serve as precursor gases, and the deposition occurs with the chamber and table at 100 °C. Although it could produce higher-quality oxide films, higher deposition temperatures negatively affect the monolayers, resulting in quenching and the appearance of dark spots on the monolayer superlattices. After every monolayer transfer step in the superlattice construction, we remove the PDMS film by storing the sample under vacuum for one day (pressure in the mbar range). Vacuum storage guarantees good and homogenous contact between the layers, further improved by heating at 70 °C for 10 minutes on a hotplate in air. Next, the sample cools down for 5 minutes. Then, the PDMS is peeled off carefully using sharp tweezers, leaving a new $WS_2$ monolayer on the superlattice. This method of PDMS removal is gentle on the TMD, as evidenced by the consistent fluorescence lifetime measurements taken before and after the process.[37] Note that during the final assembly step of the superlattice, we leave the PDMS as a superstrate for the top monolayer, which did not undergo ALD processing. The bottom monolayer, on the other hand, went through ALD processing 1, 2, and 3 times for the superlattice areas in Figure 5 with 2, 3, and 4 monolayers, respectively.

**Optical characterization**

We use a sample-scanning confocal microscope for transmission, reflection, and photoluminescence imaging and spectroscopy. The setup can also perform hyperspectral imaging by recording a spectrum at every point. For reflection and transmission measurements, we illuminate the sample with an incandescent white light source using Köhler illumination modules from the top or the bottom of the sample. For the bottom path used in transmission configuration, we use a microscope objective with adjustable cover-glass correction (Nikon CFI Plan Fluor ELWD 20x, NA = 0.45) to illuminate the sample through the supporting glass slide. The transmitted or reflected light is collected through the top objective (Nikon CFI Plan Fluor ELWD 40x, NA = 0.6), allowing for optical thickness correction of a possible PDMS top layer. After attenuation by a neutral density filter, light is coupled into a fiber with a core size of 50 μm. The signal is then guided into a spectrometer (Andor Shamrock 303i spectrograph with an Andor Newton 970 EMCCD camera cooled to -70 °C) to carry out spectroscopy or hyperspectral imaging, or to an avalanche photodiode (Micro Photon Devices, PDM50) for imaging.

For photoluminescence measurements, we use a continuous-wave laser at 532 nm (Cobolt Samba) with neutral density filters to control the power reaching the sample in the 1−100 μW range, depending on PL efficiency changes due to doping and quenching. The excitation laser is cleaned using a band-pass filter (Thorlabs, FLH532-4) and reflected towards the sample by a non-polarizing beam splitter (Chroma, 21014 Silver Non-Polarizing 50/50 bs). It is focused on the sample by the same top objective, resulting in photoluminescence emission filtered using a long-pass filter (Thorlabs, FELH0550) and collected in epifluorescence configuration through the same optical path as the reflection measurements.

**Supporting Information**

Comparison of $WS_2$ monolayer on PDMS and on glass; Transmittance, reflectance, and absorptance spectra; Transfer matrix method and the maximum absorption limit; Hyperspectral analysis of different spacers; Effect of atomic layer deposition on $WS_2$ monolayer.

**Acknowledgments**

This work was financially supported by the Netherlands Organization for Scientific Research (NWO) through an NWO START-UP grant (740.018.009). We thank Raziman T. V. and Ershad Mohammadi for stimulating discussions.

**References**

1. Woltersdorff, W. Über die optischen Konstanten dünner Metallschichten im langwelligen Ultrarot. *Zeitschrift für Phys.* **91**, 230–252 (1934).
2. Kats, M. A. & Capasso, F. Optical absorbers based on strong interference in ultra-thin films. *Laser Photonics Rev.* **10**, 735–749 (2016).

3. Li, Y. & Heinz, T. F. Two-dimensional models for the optical response of thin films. *2D Mater.* **5**, 025021 (2018).

4. Lee, S. *et al.* Achieving near-perfect light absorption in atomically thin transition metal dichalcogenides through band nesting. *Nat. Commun.* **14**, 3889 (2023).

5. Horng, J. *et al.* Perfect absorption by an atomically thin crystal. *Phys. Rev. Appl.* **14**, 024009 (2020).

6. Chaves, A. J. *et al.* Near-Unity Light Absorption in a Monolayer $WS_2$ Van der Waals Heterostructure Cavity. *Nano Lett.* **20**, 3545–3552 (2020).

7. Wang, S. *et al.* Collective Mie Exciton-Polaritons in an Atomically Thin Semiconductor. *J. Phys. Chem. C* **124**, 19196–19203 (2020).

8. Butun, S. *et al.* Quantifying Plasmon-Enhanced Light Absorption in Monolayer $WS_2$ Films. *ACS Appl. Mater. Interfaces* **9**, 15044–15051 (2017).

9. Bahauddin, S. M., Robatjazi, H. & Thomann, I. Broadband Absorption Engineering to Enhance Light Absorption in Monolayer $MoS_2$. *ACS Photonics* **3**, 853–862 (2016).

10. Huang, L. *et al.* Atomically Thin $MoS_2$ Narrowband and Broadband Light Superabsorbers. *ACS Nano* **10**, 7493–7499 (2016).

11. Wan, W. *et al.* Time-reversed lasing and interferometric control of absorption. *Science* **331**, 889–892 (2011).

12. Woo, B. H. *et al.* Dispersion Control of Excitonic Thin Films for Tailored Superabsorption in the Visible Region. *ACS Photonics* **4**, 1138–1145 (2017).

13. Bradley, M. S., Tischler, J. R. & Bulović, V. Layer-by-layer J-aggregate thin films with a peak absorption constant of $10^2$ $cm^{-1}$. *Adv. Mater.* **17**, 1881–1886 (2005).

14. Zengin, G. *et al.* Realizing strong light-matter interactions between single-nanoparticle plasmons and molecular excitons at ambient conditions. *Phys. Rev. Lett.* **114**, 1–6 (2015).

15. Eizner, E., Avayu, O., Ditcovski, R. & Ellenbogen, T. Aluminum Nanoantenna Complexes for Strong Coupling between Excitons and Localized Surface Plasmons. *Nano Lett.* **15**, 6215–6221 (2015).

16. Bellessa, J., Bonnand, C., Plenet, J. C. & Mugnier, J. Strong coupling between surface plasmons and excitons in an organic semiconductor. *Phys. Rev. Lett.* **93**, 3–6 (2004).

17. Grigorenko, A. N., Polini, M. & Novoselov, K. S. Graphene plasmonics. *Nat. Photonics* **6**, 749–758 (2012).

18. Nair, R. R. *et al.* Fine structure constant defines visual transparency of graphene. *Science* **320**, 1308 (2008).

19. Pandey, D., Xiao, S. & Wubs, M. Graphene multilayers for coherent perfect absorption: effects

of interlayer separation. *Opt. Express* **30**, 44504 (2022).

20. Papadakis, G. T. *et al.* Ultralight Angstrom-Scale Optimal Optical Reflectors. *ACS Photonics* **5**, 384–389 (2018).

21. Green, M. A., Ho-Baillie, A. & Snaith, H. J. The emergence of perovskite solar cells. *Nat. Photonics* **8**, 506–514 (2014).

22. Bernardi, M., Palummo, M. & Grossman, J. C. Extraordinary sunlight absorption and one nanometer thick photovoltaics using two-dimensional monolayer materials. *Nano Lett.* **13**, 3664–3670 (2013).

23. Datta, I. *et al.* Low-loss composite photonic platform based on 2D semiconductor monolayers. *Nat. Photonics* **14**, 256–262 (2020).

24. van de Groep, J. *et al.* Exciton resonance tuning of an atomically thin lens. *Nat. Photonics* **14**, 426–430 (2020).

25. Kang, D. H. *et al.* High-Performance Transition Metal Dichalcogenide Photodetectors Enhanced by Self-Assembled Monolayer Doping. *Adv. Funct. Mater.* **25**, 4219–4227 (2015).

26. Back, P., Zeytinoglu, S., Ijaz, A., Kroner, M. & Imamoğlu, A. Realization of an Electrically Tunable Narrow-Bandwidth Atomically Thin Mirror Using Monolayer $MoSe_2$. *Phys. Rev. Lett.* **120**, 1–5 (2018).

27. Scuri, G. *et al.* Large Excitonic Reflectivity of Monolayer $MoSe_2$ Encapsulated in Hexagonal Boron Nitride. *Phys. Rev. Lett.* **120**, 37402 (2018).

28. Gu, J. *et al.* Enhanced nonlinear interaction of polaritons via excitonic Rydberg states in monolayer $WSe_2$. *Nat. Commun.* **12**, 2269 (2021).

29. Zhao, J. *et al.* Exciton polariton interactions in Van der Waals superlattices at room temperature. *Nat. Commun.* **14**, 1512 (2023).

30. Wang, S. *et al.* Limits to Strong Coupling of Excitons in Multilayer $WS_2$ with Collective Plasmonic Resonances. *ACS Photonics* **6**, 286–293 (2019).

31. Kumar, P. *et al.* Light–matter coupling in large-area van der Waals superlattices. *Nat. Nanotechnol.* **17**, 182–189 (2022).

32. Hu, F. & Fei, Z. Recent Progress on Exciton Polaritons in Layered Transition-Metal Dichalcogenides. *Adv. Opt. Mater.* **8**, 1–16 (2020).

33. Zhou, B. *et al.* A chemical-dedoping strategy to tailor electron density in molecular-intercalated bulk monolayer $MoS_2$. *Nat. Synth.* 1–9 (2023).

34. O'Brien, K. P. *et al.* Process integration and future outlook of 2D transistors. *Nat. Commun.* **14**, 1–5 (2023).

35. O'Brien, K. P. *et al.* Advancing 2D Monolayer CMOS Through Contact, Channel and Interface

Engineering. in *IEEE International Electron Devices Meeting (IEDM)* 7–1 (2021).

36. Chung, Y.-Y. First Demonstration of GAA Monolayer-MoS$_2$ Nanosheet nFET with 410 µA/µm ID at 1V VD at 40nm Gate Length. in *IEEE International Electron Devices Meeting (IEDM)* 34–5 (IEDM, 2022).

37. Eizagirre Barker, S. *et al.* Preserving the Emission Lifetime and Efficiency of a Monolayer Semiconductor upon Transfer. *Adv. Opt. Mater.* **7**, 1900351 (2019).

38. Chernikov, A. *et al.* Exciton binding energy and nonhydrogenic Rydberg series in monolayer WS$_2$. *Phys. Rev. Lett.* **113**, 076802 (2014).

39. Sturmberg, B. C. P. *et al.* Total absorption of visible light in ultrathin weakly absorbing semiconductor gratings. *Optica* **3**, 556 (2016).

40. Vancsó, P. *et al.* The intrinsic defect structure of exfoliated MoS$_2$ single layers revealed by Scanning Tunneling Microscopy. *Sci. Rep.* **6**, 1–7 (2016).

41. Hu, Z. *et al.* Two-dimensional transition metal dichalcogenides: Interface and defect engineering. *Chem. Soc. Rev.* **47**, 3100–3128 (2018).

42. Zhang, M. *et al.* Super-resolved Optical Mapping of Reactive Sulfur-Vacancies in Two-Dimensional Transition Metal Dichalcogenides. *ACS Nano* **15**, 7168–7178 (2021).

43. Raja, A. *et al.* Dielectric disorder in two-dimensional materials. *Nat. Nanotechnol.* **14**, 832–837 (2019).

44. Stevenson, P. R. *et al.* Reversibly Tailoring Optical Constants of Monolayer Transition Metal Dichalcogenide MoS$_2$ Films: Impact of Dopant-Induced Screening from Chemical Adsorbates and Mild Film Degradation. *ACS Photonics* **8**, 1705–1717 (2021).

45. Mouri, S., Miyauchi, Y. & Matsuda, K. Tunable photoluminescence of monolayer MoS$_2$ via chemical doping. *Nano Lett.* **13**, 5944–5948 (2013).

46. Hu, P. *et al.* Control of Radiative Exciton Recombination by Charge Transfer Induced Surface Dipoles in MoS$_2$ and WS$_2$ Monolayers. *Sci. Rep.* **6**, 24105 (2016).

47. Mouri, S., Miyauchi, Y. & Matsuda, K. Chemical doping modulation of nonlinear photoluminecscence properties in monolayer MoS$_2$. *Appl. Phys. Express* **9**, 055202 (2016).

48. Liu, B. *et al.* Engineering Bandgaps of Monolayer MoS$_2$ and WS$_2$ on Fluoropolymer Substrates by Electrostatically Tuned Many-Body Effects. *Adv. Mater.* **28**, 6457–6464 (2016).

49. Castellanos-Gomez, A. *et al.* Deterministic transfer of two-dimensional materials by all-dry viscoelastic stamping. *2D Mater.* **1**, 0–8 (2014).

50. Yang, K. *et al.* Tuning electronic behaviors of WS$_2$ by molecular doping. *Mater. Today Commun.* **33**, 104226 (2022).

51. Park, S. *et al.* Demonstration of the key substrate-dependent charge transfer mechanisms

between monolayer MoS$_2$ and molecular dopants. *Commun. Phys.* **2**, 109 (2019).

52. Jing, Y., Tan, X., Zhou, Z. & Shen, P. Tuning electronic and optical properties of MoS$_2$ monolayer via molecular charge transfer. *J. Mater. Chem. A* **2**, 16892–16897 (2014).

Supporting Information

# Approaching the absorption limit with monolayer semiconductor superlattices

Sara A. Elrafei[1], Lennart M. Heijnen[1], Rasmus H. Godiksen[1], Alberto G. Curto[1,2,3*]

[1] Department of Applied Physics and Eindhoven Hendrik Casimir Institute,
Eindhoven University of Technology, Eindhoven, The Netherlands

[2] Photonics Research Group, Ghent University-imec, Ghent, Belgium

[3] Center for Nano- and Biophotonics, Ghent University, Ghent, Belgium

* Corresponding author: Alberto.Curto@UGent.be

**Contents:**

**Supporting Section S1.** Comparison of $WS_2$ monolayer on PDMS and on glass.

**Supporting Section S2.** Transmittance, reflectance, and absorptance spectra.

**Supporting Section S3.** Transfer matrix method and the maximum absorption limit.

**Supporting Section S4.** Hyperspectral analysis of different spacers.

**Supporting Section S5.** Effect of atomic layer deposition on $WS_2$ monolayer

## Supporting Section S1. Comparison of $WS_2$ monolayer on PDMS and on glass

The choice of substrate influences the properties of $WS_2$ monolayer. The substrate can introduce strain, affect charge transfer, modify the dielectric environment, and introduce surface roughness, all of which collectively impact the electronic and optical characteristics of the monolayer. In this Section, we conduct a comparative analysis of the optical properties of monolayer $WS_2$ on various substrates such as PDMS, and glass. Supporting Figure S1a (top) shows the normalized PL of the two samples. Ideally, one would compare the same monolayer on both substrates, but transferring a monolayer between substrates is challenging as it causes damage. Instead, we used separate monolayers while ensuring they had comparable quality by examining their photoluminescence lifetime, which serves as an indicator of the overall exciton quality. We see a significant increase in linewidth and a redshift of around 12 meV for the glass substrate. This shift can be attributed to the higher permittivity of glass compared to PDMS, resulting in a lower permittivity contrast with the monolayer; as a result, stronger dielectric screening for the monolayer on glass leads to a decreased binding energy and a shift of the band gap to lower energy, an effect discussed by Raja *et al*. (Nat. Nanotechnol. 14, 832–837 (2019)).

In Supporting Figure S1a (bottom), we present the transmittance of $WS_2$ monolayers deposited on PDMS and glass substrates. The A-exciton peaks are at 613 nm (2.023 eV) for PDMS and 616 nm (2.013 eV) for glass, indicating an influence of the substrate on the peak position. The monolayer on the PDMS substrate exhibited a more pronounced transmittance contrast (0.17 compared to 0.10 for glass) and a narrower linewidth (23 versus 39 meV), suggesting that PDMS is a more suitable substrate for monolayer $WS_2$. This is expected, as glass provides a less clean and less homogeneous dielectric environment for the monolayer. Finally, we also compare the quality of PDMS and glass as substrates using hyperspectral PL imaging, with part of the results summarized in the histograms in Supporting Figure S1b. The redshift in peak energy, around 9 meV on average, and the increased linewidth, by around 6 meV, of the sample on glass are both clearly visible.

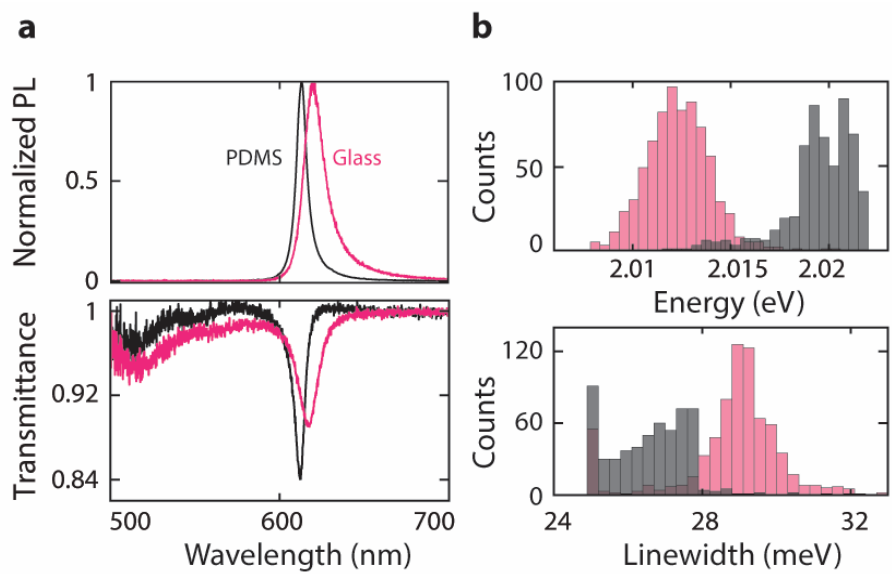

**Supporting Figure S1** | Using PDMS as a substrate for $WS_2$ monolayer results in high absorption, narrow exciton linewidth, and ensures uniform properties compared to a glass substrate **a,** Comparison of the normalized photoluminescence and transmission spectra of a $WS_2$ monolayer on glass and PDMS substrates. **b,** Comparison of histograms obtained through hyperspectral imaging of the photoluminescence of monolayers on the same substrates.

## Supporting Section S2. Transmittance, reflectance, and absorptance spectra

As part of our experiments, we measured transmittance and reflectance to determine the absorptance of three types of thin-film structures of monolayer $WS_2$, including air-spacer bilayers (or artificial bilayers), molecular-spacer bilayers, and $Al_2O_3$-spacer superlattices. We observed that reflectance played a small role in the optical response of monolayers and bilayers, which led us to refer primarily to transmission data for ease of comparison by other authors as it relies on a single measurement but serves as a good proxy for absorption. The low reflectance of monolayers and bilayers compared to peak transmittance contrast is indeed predicted by Figure 1c because the refractive index is too low to produce significant reflectance for such a small thickness. However, we show next that this simplified treatment is less applicable in the case of higher-order superlattices. We present the result of such optical measurements in Supporting Figure S2. The artificial bilayer exhibits a substantial improvement in transmission contrast compared to a monolayer, accompanied by a slight increase in the reflectance spectrum from a reflectance peak of 1% to approximately 3.7%. Following the equation $A = 1 - R - T$, we computed the corresponding absorptance spectra, which results in an increase in the absorptance peak from 17% to 27%. For the molecular-spacer bilayer, where we employed TCNQ molecules as controlled spacers, we achieved a similar absorptance close to 27%. In the case of higher-order superlattices incorporating multiple $Al_2O_3$ layers using atomic layer deposition alternating with monolayer $WS_2$, the reflectance peak reaches approximately 9% for structures with four monolayers, which yields a maximum absorptance of 31%.

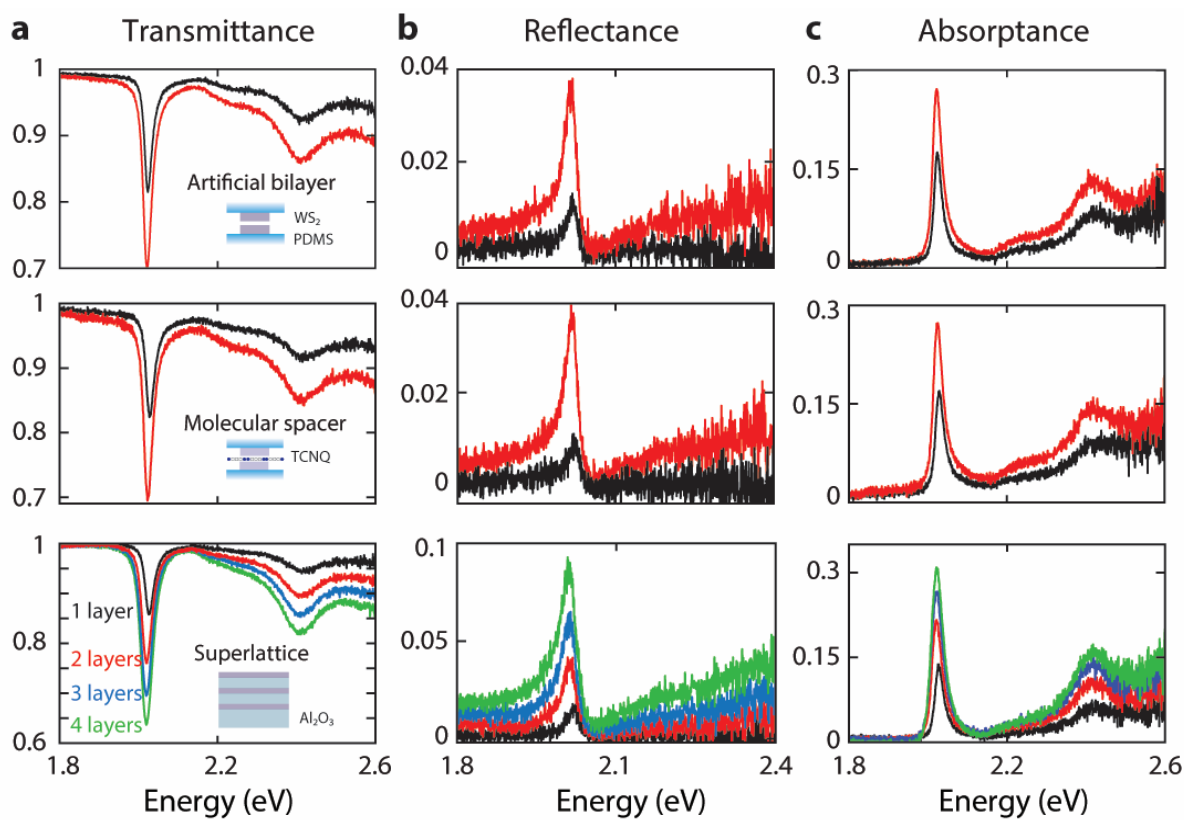

**Supporting Figure S2** | Analysis of stacks of $WS_2$ monolayers with various spacer layers: air (top), molecular (middle), and $Al_2O_3$ spacers (bottom). **a,** Transmittance. **b,** Reflectance. **c,** Absorptance.

**Supporting Section S3. Transfer matrix method and the maximum absorption limit**

We employ the transfer matrix method (TMM) to determine the transmittance, reflectance, and absorptance of artificial stacks of $WS_2$ monolayers. The surrounding medium is typically either air or PDMS, depending on the situation. We use the refractive indices of PDMS and air as 1.45 and 1, respectively. For the $WS_2$ monolayer, we use our experimental permittivity shown in Supporting Figure S3a, which was modeled using a Lorentzian model with 4 oscillators: $\varepsilon(E) = \varepsilon_{background} + \sum_{i=1}^{i=4} f_i/(E_{i,exciton}^2 - E^2 - i\gamma_i E)$, where $\varepsilon_B$ is the background permittivity, $f_i$ is the oscillator strength, $E_{i,\,exciton}$ is the resonance exciton energy, $E = \hbar\omega$ is the photon energy, and $\gamma_i$ is the linewidth of each exciton absorption band.

For each layer, we constructed a transfer matrix ($T_{ij}$) that describes how a wave passes through the interface from layer $i$ to layer $j$ within the structure. Using the Fresnel equations, we calculated the reflection ($R$) and transmission ($T$) coefficients at each interface, which are part of the transfer matrix. We then multiplied these transfer matrices sequentially to obtain the overall transfer matrix for the entire structure: $T_N = T_{1-2} \ldots\ldots T_{N-1}$. From this matrix, we derived the absorptance using $A = 1 - R - T$. These calculated coefficients and absorptance values were used to plot the optical properties of our structure as a function of wavelength, refractive index, and thickness.

Next, we evaluate the response of a thin film made of layered semiconductor material with thickness $d$ and complex refractive index $n + ik$ sandwiched between two semi-infinite media. We consider only normal incidence. The incident light has a vacuum wavelength $\lambda_0$ with its associated wavenumber $q_0 = 2\pi/\lambda_0$. The wavelength-normalized film thickness is $\delta \equiv q_0 d \ll 1$. Importantly, this thickness must be significantly smaller than the wavelength of light, both within and outside the film, to maintain a wavelength-normalized optical path length $|m\delta|$ below 1. Figure S3b depicts the absorptance for four different film thicknesses as a function of the real and imaginary parts of the refractive index ($n$ and $k$). The simulation includes thin films of thickness equivalent to a variable number of monolayers, each with the permittivity of a single monolayer and a thickness of 0.618 nm, and a choice of substrate and superstrate: air ($n = 1$), glass (1.5), and the polymer PDMS (1.42). Notably, we observe absorptance maxima approaching 50% along the dashed line where $n = k$. For a monolayer in an asymmetric environment, surrounded by PDMS on one side and air on the other, the absorption limit of 50% cannot be reached. Full encapsulation in a symmetric PDMS environment thus increases peak absorption. When the number of layers increases, the absorption maximum moves to smaller $n$ and $k$ values, which are easier to find in realistic materials. Our experimental WS$_2$ monolayer refractive index is around $5 + 5i$ close to the A-exciton peak represented by the arrows at energy 2.023 eV in the first panel. Therefore, we expect to reach the absorption limit with a stack of 8 monolayers (if they preserve their intrinsic properties after stacking).

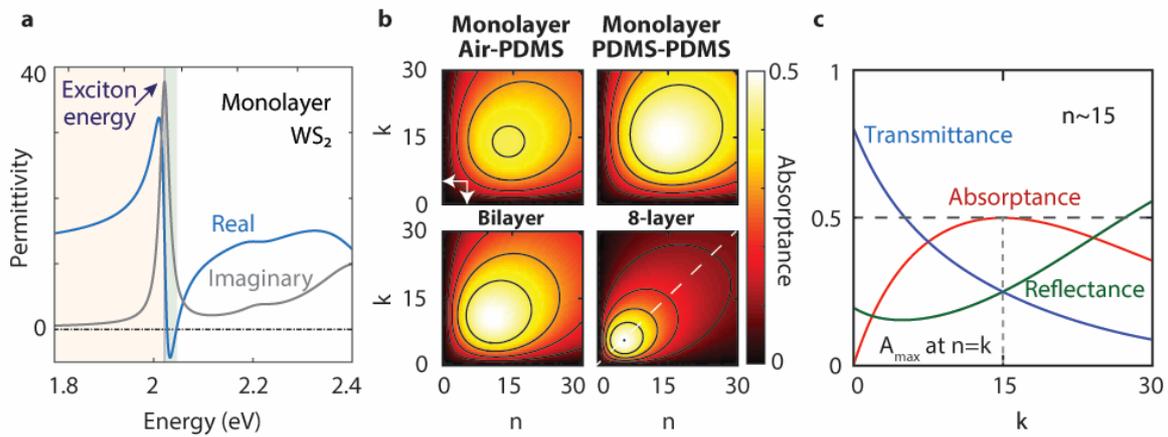

**Supporting Figure S3 |** Conditions for reaching the maximum absorption limit. **a,** In-plane permittivity of monolayer WS$_2$, derived by fitting the transmission spectrum using the transfer matrix method and a 4-Lorentzian permittivity model. **b,** Top: absorptance of a layer with thickness 0.618 nm at $\lambda = 613$ nm for varying real and imaginary parts of the refractive index $n$ and $k$, for two different superstrates (air with $n_{air}=1$, or PDMS with $n_{PDMS} = 1.42$) on a PDMS substrate. Bottom: absorptance of films with a thickness of 2 and 8 times that in the top panels, surrounded by PDMS. **c,** Calculated transmittance, reflectance, and absorptance for a film with monolayer thickness as a function of the imaginary part of the refractive index, $k$, at a constant $n$, illustrating the condition for the absorption limit at $n = k$.

**Supporting Section S4. Hyperspectral analysis of different spacers**

Hyperspectral imaging is a versatile tool for evaluating the homogeneity of monolayer $WS_2$ samples. By analyzing parameters such as PL intensity, linewidth, and peak energy, we can gain valuable insights into the uniformity and consistency of the samples. In Supporting Figure S4, we compare the spatial distribution of A-exciton peak energy for the three types of monolayer structures studied in the main text fitted from PL hyperspectral images.

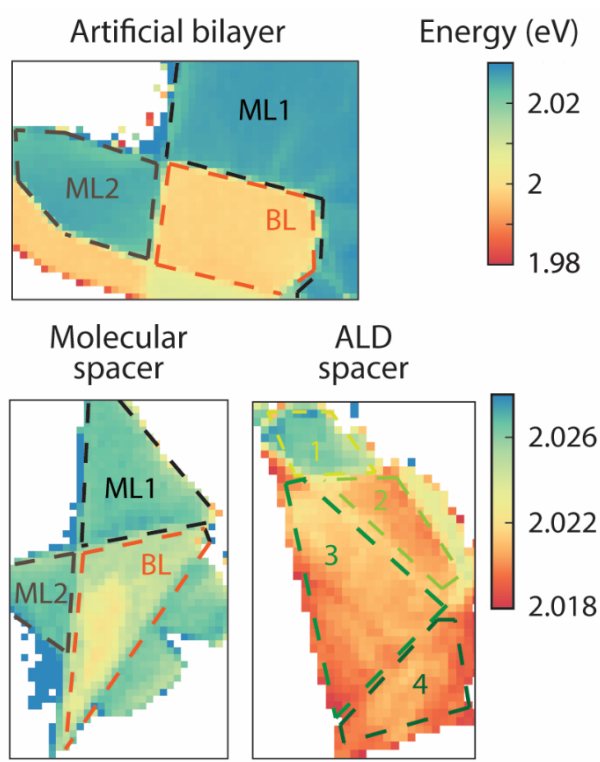

**Supporting Figure S4** | Hyperspectral imaging showing the spatial variations of A-exciton peak energy for stacks of $WS_2$ monolayers with different spacer materials.

## Supporting Section S5. Effect of atomic layer deposition on WS2 monolayer

Atomic layer deposition (ALD) enables the creation of uniform and conformal coatings. However, the direct deposition of $Al_2O_3$ onto monolayer $WS_2$ via ALD has discernible effects on its optical response, that change its photoluminescence and excitonic behavior. ALD requires solid substrates, as opposed to the polymeric PDMS substrate used elsewhere in this work. Therefore, first we use lifetime measurements to evaluate the monolayer quality before and after depositing it on a different substrate. The photoluminescence decay of $WS_2$ on a PDMS substrate is slower, indicating longer exciton lifetimes. In contrast, when we transfer a monolayer $WS_2$ on $Al_2O_3$-coated amorphous quartz, the lifetime is reduced caused by changes in the electronic environment and increased non-radiative processes (Supporting Figure S5a). On the other hand, in the ALD process, even with a low deposition temperature of 100 °C to minimize damage, the additional defects due to the ALD chemical reaction process negatively affect the quantum efficiency of the monolayer (Supporting Figure S5b).

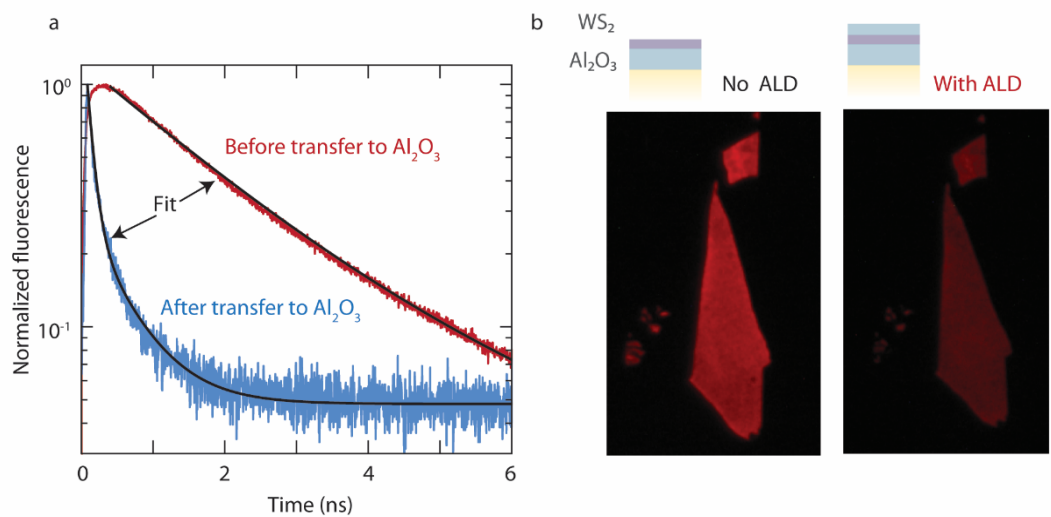

**Supporting Figure S5** | Impact of atomic layer deposition on a $WS_2$ monolayer **a,** Comparison of lifetimes before and after transfer to an $Al_2O_3$-coated substrate. **b,** Widefield PL images of an as-exfoliated monolayer $WS_2$ on $Al_2O_3$/amorphous quartz, and after depositing 2 nm of $Al_2O_3$ on it using ALD.